\documentstyle[12pt]{article}

\newcommand{\reff}[1]{(\ref{#1})}
\def\reff #1{(\ref{#1})}
\def\bra #1{\langle{#1}|}
\def\ket #1{|{#1}\rangle}

\def\beqn{\begin{eqnarray}}
\def\eeqn{\end{eqnarray}}
\def\beq{\begin{equation}}
\def\eeq{\end{equation}}
\def\phan{\vphantom{\frac12}}

\def\non{\nonumber}
\def\al{\alpha}
\def\be{\beta}
\def\ga{\gamma}
\def\Ga{\Gamma}

\def\La{\Lambda}
\def\veps{\varepsilon}
\def\D{{\cal D}}
\def\L{{\cal L}}
\def\de{\delta}
\def\cP{{\cal P}}
\def\({\left(}
\def\){\right)}
\def\l{\left\{}
\def\r{\right\}}
\author{S. M. Klishevich%
 \thanks{E-mail address: klishevich@mx.ihep.su} \\
         {\it  Institute for High Energy Physics
         } \\
         {\it Protvino, Moscow Region, 142284, Russia
         }
       }
\title{Massive Fields with Arbitrary Half-Integer Spin in Constant
Electromagnetic Field}
\begin{document}
\maketitle
\begin{abstract}
\thispagestyle{empty}
We study the interaction of gauge fields of arbitrary half-integer spins
with the homogeneous electromagnetic field. We reduce the problem of
obtaining the gauge-invariant Lagrangian and transformations of the
half-integer spin fields in the external field to an algebraic problem of
search for a set of operators with certain algebraical features using the
representation of the higher-spin fields as vectors in a pseudo-Hilbert
space. We consider such construction at linear order in the external
electromagnetic field and also present an explicit form of 
interaction Lagrangians and gauge transformations for the massive
particles of spins $\frac32$ and $\frac52$ in terms of symmetric
spin-tensor fields. The obtained result is valid for space-time of
arbitrary even dimension.
\end{abstract}
\vskip2mm
\noindent
{\it PACS number(s):} 11.10.Kk, 11.10.Lm, 11.15.-q, 11.15.Ex 
\vskip2mm
\noindent

\newpage
\setcounter{page}{1}
\section{Introduction}

At present there are a lot of different approaches to the
description of free higher-spin fields (see for, example, Refs.%
\cite{Sing-Hagen}\nocite{Fang,Fronsdal-1,Vasilev:80YF,%
Zinoviev-83,Rindani:MPL88b,}-\cite{Pashnev-1}). But the investigation of 
interactions usually faces significant difficulties. Very often the study
of the higher-spin fields ends at the free level. In this work we develop
the algebraic approach that allows us to construct an interaction 
for all the massive half-integer spin fields at once.

As is well-known, the massless fields of spins $s\ge\frac32$ do not
have the "minimal" interaction with electromagnetic field in asymptotically
flat space-time. This is related to the impossibility to
construct linear approximation in such a case \cite{Zinoviev-2}. But
the massive higher-spin fields can have the interaction 
\cite{argyres,mass_spin,spin3_1,oscB}. Therefore, we will
study the massive case only.

 In the literature the electromagnetic interaction of arbitrary spin fields
was considered at the lowest order \cite{FPT-92}. Investigating the
interactions, the authors started from the free theory of massive fields
in the conventional form \cite{Sing-Hagen}. The "minimal" introduction of
the interaction leads to contradictions. Therefore, it is necessary to
include non-minimal terms into an interaction Lagrangian. Since the massive
Lagrangian for the higher-spin fields \cite{Sing-Hagen} is not gauge
invariant, there are no restrictions on the form of the non-minimal
interaction in such approach and additional restrictions have to be imposed
in order to build a consistent theory. So, for instance, studying the
electromagnetic interaction \cite{FPT-92}, the authors have used the
requirement that tree-level scatering amplitudes must possess a smooth
$M\to0$ fixed-charge limit for any theory describing the interaction of
arbitrary-spin massive particles with photons. Under such requirement, the
amplitudes do not violate unitarity up to center-of-mass energies $E\gg
M/e$. This restriction leads to the gyromagnetic ratio $g=2$ for massive
particles of any spin. In Ref.\cite{Nappi-PR89} the authors were
investigating the electromagnetic interaction of massive spin-2 field
using the compactification of the~5 dimension gravity. But that approach
does not work (in any case in the asymptotically flat space) for the fields
of higher spins since it implies the existence of an consistent theory
of the interaction for the massless higher-spin fields. In
Ref.\cite{argyres} the way to obtain the e.m. interaction of the massive
fields of arbitrary spins has been proposed. This method is founded on the
formulation of the open bosonic string in the external constant field. But
in Ref.\cite{spin3_1} it has been shown that such approach allows one to
get the e.m. interaction of the fields for the whole string mass level
rather than for the single massive field, because the presence of the
interaction mixes the states at given mass level.

Here we go along the line of Ref.\cite{oscB} where the massive
integer-spin fields were investigated. We consider the interaction of
massive fields of arbitrary half-integer spins with the homogeneous
electromagnetic field at linear order.

We represent free state with the half-integer spin $s+\frac12$
as state~$\ket{\Psi^s}$ in a pseudo-Hilbert space\footnote{In a similar way
the representation of massless free fields was considered in Refs.%
\cite{Pashnev-1,Pashnev:MPL97} for arbitrary integer spins and in
Ref.\cite{Domeij:90IJMP} for half-integer spins.}. This space contains
the bosonic one \cite{oscB} as subspace. Coefficient functions of the
state~$\ket{\Psi^s}$ are spin-tensor fields corresponding to a particle with
spin~$s+\frac12$. In the considered Fock space we introduce a set of even
and odd operators. By means of these operators we define the gauge
transformations and the necessary constraints for the state~$\ket{\Psi^s}$.
Like the bosonic case the gauge-invariant Lagrangian has the form of the
expectation value of a Hermitian operator in state~$\ket{\Psi^s}$ but the
operator is odd in this case.

 In the considered approach the gauge invariance is a consequence of
commutation relations of the introduced operators. The introduction of the
interaction by means of the "minimal" coupling prescription induces a
change of algebraic features of the operators and, as a consequence, leads
to the loss of the gauge invariance. The problem of restoring the
invariance is reduced to the algebraic problem of search for such
modificated operators that restore the initial commutation relations. We
should note that in the massless case one cannot realize such a
construction. This relates to the fact that when the interaction is present,
the massless limit does not exist.

 In section \ref{FInter} we construct the set of the operators having the
necessary features at linear order in the external e.m. field. Besides, in
the next section we give an explicit form of the interaction Lagrangian and
the transformations for the massive spin-$\frac32$ and spin-$\frac52$ fields
in this approximation.

\section{Gauge Massive Fields with Integer Spins}
\label{Free}
Here we will briefly discuss a description of the massive fields with an
integer spins using an auxiliary Fock space.

 Let us consider the Fock space generated by creation and annihilation
operators $\bar a_\mu$, $a_\mu$ with a Lorentz vector index and by the
scalar ones $\bar b$ and $b$. These operators have the commutation relations
\begin{eqnarray}
\label{H-alg}
\left[a_\mu,\bar a_\nu\right]&=&g_{\mu\nu}, \quad a_\mu^{\dag} = \bar
a_\mu,
\non\\
\left[b,\bar b\right]&=&1,\quad b^{\dag}=\bar b.
\end{eqnarray}
where $g_{\mu\nu}$ is the metric tensor on space-time ${\cal M}_D$ of
arbitrary dimension $D$  with the signature $\|g_{\mu\nu}\|={\rm
diag}(-1,1,1,...,1)$. Since the metric is indefinite, the Fock space, which
realizes the representation of the Heisenberg algebra \reff{H-alg}, is
pseudo-Hilbert. Therefore, to exclude the states with negative norms, we
have to impose additional conditions on the physical states.
 
We will consider the states of type
\begin{equation}
\label{FockM}
\ket{\Phi^s}=\sum\limits_{n=0}^s\Phi_{\mu_1\dots\mu_{n}}(x) 
\bar b^{s-n}\prod_{i=1}^n\bar a_{\mu_i}\ket{0},
\end{equation}
where $\ket{0}$ is the usual Fock vacuum. The coefficient functions
$\Phi_{\mu_1\dots\mu_n}(x)$ are symmetric tensor fields of rank $n$
on~${\cal M}_\D$. At $s\to\infty$ such states span the whole Fock space.

In order to properly describe the physical state with spin $s$ by vector
\reff{FockM}, we should impose a restriction on this state. For that, in the
pseudo-Hilbert space we introduce the following operators:
\begin{eqnarray}
\label{L-operator0}
&L_1=p_\mu a^\mu + mb,\quad L_{-1}=L_1^{\dag},&
\non\\
&L_2=\frac12\(a_\mu a^\mu + b^2\),\quad L_{-2}=L_2^{\dag},&
\\\non
&\quad L_0=p^2 + m^2.&
\end{eqnarray}
Here $p_\mu=i\partial_\mu$ is the momentum operator, which acts on the
coefficient functions.

Operators \reff{L-operator0} satisfy the following commutation relations:
\begin{equation}
\label{L-algebra0}
 \begin{array}{rclrcl}
 \left[L_1,L_{-2}\right]&=&L_{-1},
&\quad\left[L_1,L_2\right]&=&0,\\
 \left[L_2,L_{-2}\right]&=& N + \frac{D+1}{2},
&\quad\left[L_0,L_n\right]&=&0,\\
 \left[L_1,L_{-1}\right]&=&L_0,
&\quad\left[N,L_n\right]&=&{}-nL_n,\quad n=0,\pm 1,\pm 2,
 \end{array}
\end{equation}
 where $N=\bar a_\mu a^\mu + \bar bb$. Vectors of type \reff{FockM} are
eigenvector of this operator, i.e. $N$ defines the spin of the state
$$
N\ket{\Phi^s} =s\ket{\Phi^s}.
$$
Let us impose the following condition on state \reff{FockM}
\begin{equation}
\label{Trace-L}
(L_2)^2\ket{\Phi^s}=0.
\end{equation}
This corresponds to the usual condition that the tensor fields describing
massive (massless) higher-spin particles \cite{mass_spin,oscB} is
twice-traceless.

In order to avoid the presence of the redundant states, we must also have
the gauge transformations for state \reff{FockM} in the form 
\begin{equation}
\label{Gauge-L}
\de\ket{\Phi^s}=L_{-1}\ket{\La^{s-1}}.
\end{equation}
Here the gauge Fock vector
\beq\label{GaugeStateB}
\ket{\La^{s-1}}=\sum\limits_{n=0}^{s-1}\La_{\mu_1...\mu_n}
\bar b^{s-n-1}\prod_{i=1}^n\bar a_{\mu_i}\ket{0},
\eeq
is eigenvector of operator $N$ with eigenvalue $s-1$. This vector
satisfies the condition
\begin{equation}
\label{Gtraceless}
L_2\ket{\La}=0.
\end{equation}
It is easy to verify that these relations define the usual gauge
transformations for the coefficient functions.
 
The gauge Lagrangian for the massive fields with the integer spin can be
written as the expectation value of a Hermitian operator in
state~\reff{FockM}
\begin{equation}
\label{Llagr}
\L_s=\bra{\Phi^s}\L(L)\ket{\Phi^s}, \quad \bra{\Phi^s}=\ket{\Phi^s}^{\dag},
\end{equation}
where
\begin{eqnarray}
\label{L-action}
\L(L)&=&\frac12L_0-\frac12L_{-1}L_1-L_{-2}L_0L_2
-\frac12L_{-2}L_{-1}L_1L_2
\non\\&&{}
+\frac12\left\{L_{-2}L_1L_1+\mbox{h.c.}\right\}.
\end{eqnarray}
Lagrangian \reff{Llagr} is invariant under transformations \reff{Gauge-L} as
a consequence of \reff{Gtraceless} and of the relation
$$
\L(L)L_{-1}\sim(...)L_2.
$$

In the free case one can regard this construction as the dimensional
reduction ${\cal M}_{D+1}\to{{\cal M}_D\otimes S^1}$ of the massless theory
with the radius of the sphere $R\sim1/m$ (refer also to Refs.%
\cite{Pashnev-1,Pashnev:MPL97}). We should note that this statement is
not valid when an interaction is present because in this case the terms
corresponding to the interaction are proportional to inverse degrees of
the mass parameter. For example, operator $L_1$ deformed in the presence of
the constant electromagnetic field \cite{oscB} has the following form at
linear order
\begin{eqnarray}
\label{L^(1)}
L_1^{(1)}&=&\non
\frac1m\( 1 - d_2\)\(\bar\al F\al\)\be
+\frac1{m^2}\biggl\{
\(\cP F\al\)\Biggl(d_1\(\frac12-\bar\be\be\)e^{-2\bar\be\be}
\\&&
 + d_2\(\frac12+\bar\be\be\)\Biggr)
+\(\bar\al F\cP\)\(d_1e^{-2\bar\be\be} + d_2\)b^2\biggr\},
\end{eqnarray}
where $\al_\mu$ and $\be$ are normal symbols of operators $a_\mu$ and $b$,
correspondingly. Obviously, in linear approximation action \reff{Llagr}
contains the terms proportional to inverse degrees of the mass parameter $m$
as well. Thereby one cannot perform the smooth transition to the massless
case in the presence of the interaction.

\section{Massive Half-integer spin fields}\label{SecHalf}
In this section we develop a similar construction for the gauge fields
with half-integer spins. 

The massless fermionic gauge field with spin $s+1/2$ are usually described
by means of symmetric spin-tensor $\Psi_{\mu_1\dots\mu_s}$.

Henceforth we will use the following definition:
$$
\Psi'=\ga^\mu\Psi_{\mu\mu_2\dots\mu_s}.
$$

The massless gauge field satisfies the condition
\begin{equation}
\label{Trace1}
\Psi'''=0.
\end{equation}
The gauge transformation for this field is
\begin{equation}
\label{Gauge-0}
\de\Psi_{\mu_1\dots\mu_s}=\partial_{(\mu_1}\xi_{\mu_2\dots\mu_s)},
\end{equation}
where $\xi$ is a fermionic gauge parameter, which obeys the condition
\begin{equation}
\label{Trace2}
\xi'=0.
\end{equation}

Transformation \reff{Gauge-0} and conditions \reff{Trace1}, \reff{Trace2}
unambiguously determine the Lagrangian for the free field up to surface
terms
\begin{eqnarray}
\label{LF-free}
\L^F_{s+1/2}&=&i
\left\{
   \bar\psi\cdot\hat\partial\psi+s\bar\psi'\cdot\hat\partial\psi'
   -\frac14s(s-1)\bar\psi''\cdot\hat\partial\psi''
\right.
\non\\&&
\left.{}
-2s\bar\psi'\cdot(\partial\cdot\psi)
+s(s-1)\bar\psi''\cdot(\partial\cdot\psi')
\phan\right\}.
\end{eqnarray}
One can easily see that the Lagrangian does not depend on space-time
dimensionality explicitly. 

In Ref.\cite{Rindani:MPL88f} the authors derived a gauge description of
the massive free half-integer spin fields by the dimensional reduction of
action \reff{LF-free}. Our way to obtain the action for the fermionic
fields resembles this construction in non-interacting case but
we use an auxiliary Fock space for that (for the massless fields see also
Ref.\cite{Domeij:90IJMP,Domeij:90}). 

In order to describe the massive fermionic fields in a Fock space, we
enlarge the pseudo-Hilbert space generated by operators \reff{H-alg} by
means of the anticommuting operators that obey the following relations
\beq
\label{Gamma}
\l\Ga_\mu,\Ga_\nu\r = 2g_{\mu\nu},\qquad 
\l\Ga_\mu,\bar\Ga\r = 0,\qquad \bar\Ga^2 = 1.
\eeq
We will consider these operators as the Hermitian ones. We will also 
consider space-time of even dimensionality only. This implies that the
operator $\bar\Ga$ is not independent
$$
\bar\Ga\sim \Ga_0\Ga_1...\Ga_{D-1}.
$$

To realize a representation of algebra \reff{Gamma}, we introduce
a spinor vacuum vector. This vector transforms as  spinor under the Lorentz
transformations and, therefore, it carries spinor index
\beq
 \ket{0}_F=\ket{0}_\al
\eeq
Let us define the action of operators \reff{Gamma} on the spinor vacuum by
\beqn\label{Grepr}
\non
	\Ga_\mu\ket{0}_\al &=& (\bar\ga\ga_\mu)_{\be\al}\ket{0}_\be,
\\
	\bar\Ga\ket{0}_\al &=& \bar\ga_{\be\al}\ket{0}_\be.
\eeqn
Here, matrixes $\ga_\mu$ and $\bar\ga$ have the usual properties 
\beqn\label{UsualClif}
\non &
\{\ga_\mu,\ga_\nu\} ={}-2g_{\mu\nu},\qquad  
\{\ga_\mu,\bar\ga\}=0,\qquad 
\ga_0\ga_\mu^{\dag}\ga_0 =\ga_\mu,
& \\ &
\bar\ga = (-1)^{\frac14(D-2)}\ga_0\ga_1...\ga_{D-1},\qquad
 \bar\ga^{\dag} = \bar\ga.
&
\eeqn
We also define the dual vacuum vector and the scalar product in the
following way:
\beq\label{VacDef}
	{}_\al\langle0|0\rangle_\be = (\ga_0\bar\ga)_{\al\be},\qquad
{}_\al\bra{0}=\ket{0}_\al^{\dag}.
\eeq

From definitions (\ref{Grepr}-\ref{VacDef}) it is not difficult
to check that the operators $\Ga_\mu$ and $\bar\Ga$ are Hermitian indeed.

The whole vacuum vector is the tensor product of the bosonic and fermionic
vacua
$$
\ket{0}=\ket{0}_B\otimes\ket{0}_F.
$$

By means of operators \reff{Gamma} we define the following odd operators:
\beqn\label{FOp}
\non
F_1&=&\frac12\(\hat a + \bar\Ga b\),
\\
F_0&=&\hat p + m\bar\Ga,
\eeqn
where the notation $\hat A = A_\mu\Ga^\mu$ has been used.

Using relations \reff{H-alg}, \reff{L-operator0}, \reff{Gamma} and
\reff{FOp}, one can easily verify that the operators $L_i$ and $F_i$ have
the commutation relations:
\beqn\label{Falg}
 &\non
\l F_1,\bar F_0\r = L_1, \quad
F_0^{{}2}=L_0, \quad
F_1^{{}2} =\frac12L_2
&\\&
\left[L_1,\bar F_1\right] = \frac12 F_0, \quad
\left[L_2,\bar F_1\right] = F_1, \quad
\left[F_1,L_0\right] = 0. &
\eeqn
One can regard relations \reff{L-algebra0} and \reff{Falg} as a
finite-dimensional truncation of the infinite superconformal algebra
\cite{Strings}.

We will consider the fermionic state with spin $s+\frac12$ in the form of a
vector of type
\beq\label{Fstate}
\ket{\Psi^s}=\sum\limits_{n=0}^s\Psi_{\mu_1\dots\mu_{n}}(x)
\bar b^{s-n}\prod_{i=1}^n\bar a_{\mu_i}\ket{0}.
\eeq
Here we imply that coefficient functions $\Psi_{\mu_1\dots\mu_{n}}(x)$ are
spin-tensor fields on ${\cal M}_D$, i.e. they have a spinor index, which
contracts with the index of the fermionic vacuum. Henceforth we will
suppress the spinor index and will assume $m=1$ as well.

Vector \reff{Fstate} contains an redundant states like the bosonic case.
To eliminate them we impose the condition
\beq\label{F1^3}
 \(F_1\)^3\ket{\Psi^s} = 0
\eeq
and require the presence of the invariance under the gauge transformation
\beq\label{Trans0}
\de\ket{\Psi^s} = \bar L_1\ket{\xi^{s-1}},
\eeq
where the gauge vector
\beq\non
\ket{\xi^{s-1}}=\sum\limits_{n=0}^{s-1}\xi_{\mu_1...\mu_n}
\bar b^{s-n-1}\prod_{i=1}^n\bar a_{\mu_i}\ket{0}
\eeq
obeys condition
\beq\label{xi'}
 F_1\ket{\xi}=0.
\eeq

In terms of the coefficient functions relations \reff{F1^3}-\reff{xi'} 
similar to \reff{Trace1}-\reff{Trace2}, correspondingly, after the
dimensional
reduction up to some redefinitions \cite{Rindani:MPL88f}. In limit $m\to0$
they restore its usual form. We would like to note that this transition
is smooth, i.e. the number of physical degree of freedom does not change in
contrast to the conventional non-gauge formulation \cite{Sing-Hagen}.

Like the bosonic case we will search for the Lagrangians for the massive
half-integer spin fields in the form of the expectation value of a Hermitian
operator\footnote{Such an operator must be odd in the fermionic case.}{ }
in state~\reff{Fstate}
\beq\label{Faverage}
\L =\bra{\Psi^s}\L(F,L)\ket{\Psi^s}.
\eeq
The operator $\L(F,L)$ is defined by gauge transformation \reff{Trans0} and
conditions \reff{F1^3}, \reff{xi'} unambiguously and have the form
\beq\label{FAction}
\L(F,L)= F_0 + 4\bar F_1F_0F_1 - \bar L_2F_0L_2
	- 2\(\bar L_1F_1 - \bar L_2F_1L_1 + h.c.\).
\eeq
Having calculated expectation \reff{Faverage} we shall obtain the
Lagrangians for the fermions in terms of the spin-tensor fields. This
corresponds to the dimensional reduction of action \reff{LF-free} in
space-time with odd dimensionality $D+1$. Similar to the bosonic case we
have this correspondence only in the non-interacting case.

\section{Electromagnetic Interaction of Massive Half-Integer Spin Fields}
\label{FInter}

In this section we construct the interaction between the fermionic fields
with arbitrary spins and the homogeneous e.m. field in linear approximation
by the gauge-invariant manner.

We introduce the interaction by means of the "minimal" coupling
prescription, i.e. we replace the usual momentum operators by the
$U(1)$-covariant ones: $p_\mu\to \cP_\mu$. The commutator of the covariant
momenta defines the electromagnetic field strength
\begin{equation}
\left[\cP_\mu,\cP_\nu\right]=F_{\mu\nu}.
\end{equation}
For convenience we have included the imaginary unit and the coupling
constant into the definition of the strength tensor.

In the definition of operators \reff{L-operator0} and \reff{FOp} we replace
the usual momenta by the covariant ones as well. As a result the operators
cease to obey algebra \reff{L-algebra0}, \reff{Falg} and, correspondingly,
Lagrangian \reff{Faverage} loses the invariance under transformations
\reff{Trans0}. 

In order to restore the gauge invariance, we do not need to restore the
whole algebra \reff{L-algebra0} and \reff{Falg}. It is sufficient to ensure
the existence of the following relations:
\beqn\label{need}
\non
[L_1,L_{-1}] = L_0, & [L_2,L_{-1}] = L_1, & F_0^2 = L_0,
\\{}
[L_1,F_{-1}] =\frac12F_0, &[L_0,F_1] = 0.&
\eeqn

To restore the relations, we represent operators \reff{L-operator0}
and \reff{FOp} as normal ordered functions of the bosonic
creation-annihilation operators, the odd ones $\Ga_\mu, \bar\Ga$ as well as
of the
electromagnetic field, i.e.
\beqn\non
L_i&=&L_i\(\bar a_\mu,\bar b,a_\mu,b,\Ga_\mu,\bar\Ga,F_{\mu\nu}\),
\\
F_i&=&F_i\(\bar a_\mu,\bar b,a_\mu,b,\Ga_\mu,\bar\Ga,F_{\mu\nu}\).
\eeqn
The exact form of these operators is defined by the condition of 
restoring  commutation relations \reff{need}. We should note that it is
sufficient to define the form of operators $L_1$, $L_2$, and $F_1$, since
the others can be determined from \reff{need}. 

Since we consider the deformation\footnote{The deformation means that we
have passed to the extended universal enveloping algebra of the Heisengerg
one.}{ } of the operators we should take into account an arbitrariness in
the definition of the operators $a$, $b$, and $\Ga$ which have also
appeared. Besides, in the right-hand side of \reff{H-alg} and \reff{Gamma}
we should
admit the presence of arbitrary operator functions depending on $a$, $b$,
$\Ga$, and $F_{\mu\nu}$. In this, one must require
the condition that the deformed operators would not break the Jacobi
identities and they restore
the initial algebra in limit $F_{\mu\nu}\to 0$. The subsequent analysis
shows that the Jacobi identities and the arbitrariness in the definition of
operators $a$, $b$, and $\Ga$ allow one to restore initial algebra
\reff{H-alg} and \reff{Gamma} at linear order. This means that we can
consider the deformation of operators $L_i$ and $F_i$ only.

We shall search for the operators $L_1$, $L_2$, and $F_1$ as series in the
strength tensor of e.m. field which is equivalent to the expansion in
coupling constant.

Let us consider linear approximation.

Operator $L_1$ should be no higher than linear in operator $\cP_\mu$, since
the presence of a higher number of these operator changes the type of gauge
transformations \reff{Trans0} and, as a consequence, changes the number of
the physical degrees of freedom. Therefore, at this order we can search for
them in the form
\beqn\label{L1anz}
\non
L_1^{(1)}&=&\(\bar aFa\)h_0(\bar b,b)b + \(\cP Fa\)h_1(\bar b,b)
+ \(\bar aF\cP\)h_2(\bar b,b)b^2
 + \hat F h_3(\bar b,b)b
\\&&
	+ (aF\Ga)\bar\Ga h_4(\bar b,b)
	+ (\bar aF\Ga)\bar\Ga h_5(\bar b,b)b^2
	+ (\cP F\Ga)\bar\Ga h_6(\bar b,b)b,
\eeqn
where $\hat F=(\Ga F\Ga)$. At the same time the operators $L_2$ and $F_1$
can not depend on the momentum operators at all, since conditions
\reff{Trace-L} and \reff{F1^3} must define purely algebraic constraints on
the coefficient functions. Therefore, at this order we can choose the
operators in the following form: 
\beqn\label{L2F1anz}
\non
L_2^{(1)}&=& \(\bar aFa\)h_7(\bar b,b)b^2
    + \hat F h_8(\bar b,b)b^2
	+ (aF\Ga)\bar\Ga h_9(\bar b,b)b 
\\\non &&
	+ (\bar aF\Ga)\bar\Ga h_{10}(\bar b,b)b^3,
\\\non
F_1^{(1)}&=& (aF\Ga) h_{11}(\bar b,b)
   + (\bar aF\Ga) h_{12}(\bar b,b)b^2
   + (\bar aFa)\bar\Ga h_{13}(\bar b,b)b
\\&&
   + \hat F \bar\Ga h_{14}(\bar b,b)b.
\eeqn
Here $h_i(\bar b,b)$ are normal ordered operator functions of type
$$
h_i(\bar b,b)=\sum\limits_{n=0}^\infty h_n^i \bar b^n b^n,
$$
where $h^i_n$ are arbitrary real coefficients. We consider the
real coefficients only, since the operators with purely imaginary
coefficients do not give any contribution to the "minimal" interaction.

Let us define the particular form of functions $h_i$ from the condition
restoring commutation relations \reff{need} by operators
\reff{L1anz} and \reff{L2F1anz}. 

Having calculated \reff{need} and passing to normal symbols of the
crea\-tion and anni\-hi\-la\-tion operators, we get a system of linear
differential equations of the second order in functions $h_i(x)$, where
$x=\bar\be\be$ and $\be$ is the normal symbol of operator $b$. The number of
the equations is 36 for 20 unknowns. We will not represent these equations
here.

Of course, the system of linear differential equations is overdetermined but
nevertheless it is solvable. Resolving these equations we obtain as a result
\beqn\label{Fres}
\non
 L_2^{(1)}&=&{}-2c_2\(\Ga F\al\)\bar\Ga e^{-2\bar\be\be}\be,
\\\phan\non
L_1^{(1)}&=& \(1+2c_1\)\(\bar \al F\al\)\be
	- c_1\(\cP F\al\)\(1+2\bar\be\be\)
	+ 2c_1\(\bar \al F\cP\)\be^2
\\\phan\non&&{}
	+ \(\Ga F\al\)\bar\Ga\({\frac14}
	 + c_2\(1-2\bar\be\be\)e^{-2\bar\be\be}\)
	+ \(\cP F\Ga\)\bar\Ga\(c_1 + c_2e^{-2\bar\be\be}\)\be
\\\non\phan&&{}
	- 2c_2\(\bar \al F\Ga\)\bar\Ga e^{-2\bar\be\be}\be^2
	+ {\frac12}\({\frac14} + c_1\)\hat F\be,
\\\non&&
\\\phan\non
F_1^{(1)}&=& c_2\(\bar \al F\al\)\bar\Ga e^{-2\bar\be\be}\be
	+ {\frac12}c_2\hat F\bar\Ga e^{-2\bar\be\be}\be
	- {\frac12}c_2\(\Ga F\al\)e^{-2\bar\be\be},
\\\phan\non
F_0^{(1)} &=& \({\frac12} + 2c_1\)\(\bar \al F\al\)\bar\Ga
	- {\frac12}\({\frac14} - c_1\)\hat F\bar\Ga
	+ \biggl\{ \(\Ga F\al\)\bar \be\({\frac12}- 2c_2e^{-2\bar\be\be}\)
\\\phan&&{}
 - 2\(\cP F\al\)\bar\Ga\(c_1 - c_2 e^{-2\bar\be\be}\) + h.c.\biggr\}.
\eeqn

Thus, we have obtained the general form of the operators $L_i$ and $F_i$
that satisfy relations \reff{need} in linear approximation. This means that
Lagrangian \reff{FAction} is invariant under transformations
\reff{Trans0} at this order. One can see that the solution obtained has the
two-parametric arbitrariness.

In conclusion of this section we note that it is possible to restore whole
algebra \reff{L-algebra0}, \reff{Falg} if one sets $c_2=0$.

\section{Examples}

Here we apply the general result of the previous section to the
description of the electromagnetic interaction of massive fields with spins
$\frac32$ and $\frac52$.

\subparagraph{Spin $\frac32$.} The massive spin-$\frac32$ field is the
simplest one among the other gauge fermionic fields. Besides, interactions
of this field are the best studied in the literature Therefore, it is useful
to represent the interaction Lagrangian that one can derive by our approach.

The state that corresponds to the massive spin-$\frac32$ field is
\beq\label{State3/2}
\ket{3/2} =\((\chi\cdot\bar a) + \eta\bar b\)\ket{0},
\eeq
 while the gauge vector is defined as
$$
\ket{\La,3/2}=\xi\ket{0}.
$$
Having calculated expectation value \reff{Faverage} in this state,
we easily derive the free Lagrangian for the massive field with spin
$\frac32$ 
\beqn\label{L3/2}
\non
 \L_{3/2}^{(0)} &=& \bar\chi\cdot\(\hat p + 1\)\chi
	+ \bar\eta\(\hat p + 1\)
	- \(\bar\chi'+\bar\eta\)\(\hat p - 1\)\(\chi' + \eta\)
\\&&
	-\l\(\(\bar\chi\cdot p\) + \bar\eta\)\(\chi' + \eta\) + h.c.\r.
\eeqn
Here the "hat" denotes the contraction of vector index with the usual
$\ga$-matrixes. The kinetic terms of \reff{L3/2} are
non-diagonal\footnote{This is usual situation for the dimensional
reduction.}{ } but one can diagonalize them by the field redefinition of
type $\chi\to\chi + \ga\eta$. 

Then, from \reff{Trans0} we obtain the following gauge transformations for
the coefficient functions of state \reff{State3/2}
\begin{eqnarray*}
\non
\de_0\chi_\mu &=& p_\mu\xi,
\\
\de_0\eta &=& \xi.
\end{eqnarray*}
One can easily see from these transformations that field $\eta$ can be
completely gauged away. Having performed this, we obtain the usual
Rarita-Schwinger action \cite{Rarita}. for the massive field with spin
$\frac32$.

As next step we calculate the Lagrangian of the interaction and the
transformations in linear approximation. For that we have to modify
operators $F_i$ and $L_i$ according to solution \reff{Fres} and keep only
linear in the strength of the e.m. field terms in operator \reff{FAction}.
Then, having computed the expectation value of this deformed operator in
state~\reff{State3/2} one derives the interaction Lagrangian
\begin{eqnarray*}
 \L_{3/2}^{(1)} &=& \non
	\(\frac12 + 2c_1\) \(\bar\chi F\chi\)
	+ \frac12\(\frac14 - c_2\)\(\bar\chi\hat F\chi
	+ \bar\chi'\hat F\chi'\)
	+ \frac14\bar\eta\hat F \eta
\\\phan\non&&
	+ \biggl\{
	  c_2\(\(\bar\chi\cdot\cP\) - \bar\chi' - \bar\eta\)
		\(\(\ga F\chi\) + \hat F\eta\)
	- (c_1 - 2c_2)\bar\eta\(\cP F\chi\)
\\\phan\non&&
	+ \bar\chi'\(c_1\(\cP F\chi\)
	+ \(c_1 + c_2\)\(\cP F\ga\)\eta \)
	+ \frac34\bar\eta\(\ga F\chi\)
\\\phan&&
	+ \(\frac14 + c_2\)\bar\chi'\hat F\eta
	+ \frac14 \bar\chi'\(\ga F\chi\)
 + h.c. \biggr\}.
\end{eqnarray*}
Applying \reff{Trans0} to state \reff{State3/2} we obtain the following
expression for the gauge transformation
\beqn\label{Trans3/2}
\non\phan
\de_1\chi_\mu &=& c_1\(\cP F\)_\mu\xi
	 + \({\frac14} + c_2\)\(F\ga\)_\mu\xi,
\\\phan
\de_1\eta &=&\(c_1 + c_2\)\(\ga F\cP\)\xi
 - {\frac12}\({\frac14} + c_1\)\hat F\xi.
\eeqn
Obviously, constraints \reff{F1^3} and \reff{xi'} are trivial for the case
of the massive spin-$\frac32$ field.

One can notice that the Lagrangian and the transformation formally contain
one derivative more then in the supergravity theories with spontaneous
breaking.

It would be interesting to investigate the causality of the obtained
construction. Such an exploration will be made elsewhere.

\subparagraph{Spin $\frac52$.} Now we pass to the description of the
massive field with spin~$\frac52$. This case is less studied and, therefore,
it is even more interesting.

A state describing the massive field with spin $\frac52$ has the form:
\beq\label{State5/2}
\ket{s=5/2} = \((\bar a\cdot\Psi\cdot\bar a) + (\chi\cdot\bar a)\bar b
	 + \eta\bar b^2\)\ket{0},
\eeq
where $\Psi_{\mu\nu}$ is a symmetric spin-tensor.
At the same time the corresponding gauge vector is
\beq\label{GState5/2}
\ket{\La,5/2}=\((\xi\cdot\bar a) + \veps\bar b\)\ket{0}.
\eeq

Constraint \reff{F1^3} is still trivial for state \reff{State5/2}, while
\reff{xi'} imposes on the coefficient functions of \reff{GState5/2} the
following restriction:
\beq\label{xi'5/2}
\xi'+\veps=0.
\eeq
We discard $\veps$ by means of this relation.

Having calculated expectation value \reff{Faverage} in
state~\reff{State5/2}, we derive the Lagrangian describing the propagation
of the free massive
spin-$\frac52$ field
\beqn
\L_{5/2}^{(0)} &=&\non
2 \bar\Psi\(\hat p+1\)\Psi + \bar\chi\(\hat p+1\)\chi
 + 2 \bar\eta\(\hat p+1\) \eta
\\\non&&
-\(2 \bar\Psi'+\bar\chi\)\(\hat p-1\) \(2\Psi' + \chi\)
 - \(\bar\chi' + 2 \bar\eta\)\(\hat p - 1\)\(\chi' +  2\eta\)
\\\non&&
 - \(\bar\Psi'' - \bar\eta\)\(\hat p + 1\)\(\Psi'' - \eta\)
\\\non&&
+\biggl\{
\(2\(\bar\Psi\cdot p\) + \bar\chi\)\(2\Psi' + \chi\)
+\(\(\bar\chi\cdot p\) + 2 \bar\eta\) \(\chi' + 2 \eta\)
\\&&
+\(\bar\Psi'' - \bar\eta\)
\(2\(p\cdot\Psi'\) + \(p\cdot\chi\) + \chi'+2 \eta\)
+ h.c.\biggr\}.
\eeqn
And having computed \reff{Trans0} in proper way, we obtain the free
gauge transformations
\begin{eqnarray*}
\de\Psi_{\mu\nu} &=& p_{(\mu}\xi_{\nu)},
\\
\de\chi_\mu &=&{} - p_\mu\xi' + \xi_\mu,
\\
\de\eta &=&{} - \xi'.
\end{eqnarray*}
The same result can be obtained up to insignificant redefinitions by the
dimensional reduction \cite{Sivakumar-PR85a}.

Now let us pass to the case when the interaction is present.

According to the general scheme we have to use the
deformed operators. Retaining the linear in the strength
terms in \reff{Faverage} we arrive at the result
\beqn\non
\L_{5/2}^{(1)}&=&
	 \(\frac14 - c_1\)\(\bar\Psi\hat F\Psi
		 + \frac12\bar\chi\hat F\chi + \bar\eta\hat F\eta\)
	+ 2\(1+ 4c_1\)\bar\Psi' F\Psi'
\\\non&&{}
	+ \(\frac12 - 2c_1\)\bar\Psi'\hat F\Psi'
	- \frac18\(1 - 4c_1\)\(\bar\Psi'' - \bar\eta\)\hat F
	\(\Psi'' - \eta\)
\\\phan\non&&{}
	+ \(\frac12 + c_1\)\bar\chi F\chi
	+ \frac12\(\frac14 - c_1\)
	\(\bar\chi' + 2 \bar\eta\)\hat F\(\chi' + 2\eta\)
\\\phan\non&&{}
  + \biggl\{
	 (1-4c_2)\bar\chi\(\ga F\Psi\)
	- 4(c_1-c_2)\bar\chi\(\cP F\Psi\)
\\\phan\non&&{}
	+ \(\bar\chi' + 2\bar\eta\)\((1-4c_2)\(\ga F\Psi'\)
		 - 2(c_1-c_2)\(\cP F\Psi'\)\)
\\\phan\non&&{}
	+ 2c_1\(\bar\Psi'' - \eta\)(\hat \cP + 1)\(\ga F\chi\)
	+ \(\frac12+2c_2\)\bar\Psi'\(\ga F\Psi\)
\\\phan\non&&{}
	+ 2c_2\bar\Psi'(\hat \cP-1)\(\(\ga F\Psi\) - 2\(F\chi\) -\hat F\chi\)
\\\phan\non&&{}
	+ 2c_2\(2\(\bar\Psi\cdot\cP\)+\bar\chi\)
		\(\(\ga F\Psi\) - \(F\chi\) - \frac12\hat F\chi\)
\\\phan\non&&{}
	+ 2(c_1+c_2)\bar\Psi'\(\cP F\ga\)\chi
	- 2(1+2c_1)\(\bar\Psi F\chi\)
\\\phan\non&&{}
	+ \(\frac14+c_1\)\bar\Psi'\hat F\chi
	- 8c_1\(\bar\Psi' F\cP\)\eta
	- 2c_2\(\bar\Psi' F\ga\)\eta
\\\phan\non&&{}
	+ 2c_2\(2\(\bar\Psi\cdot\cP\)+\bar\chi\)
		\(\(\ga F\Psi\)-\(F\chi\) +\frac12\hat F\chi\)
\\\phan\non&&{}
	+ (\bar\Psi''-\bar\eta)\biggl(
		2c_2(\ga F(\cP\cdot\Psi))+3c_1(\cP F\chi)+c_2\hat F(\cP\cdot\chi)
\\\phan\non&&{}
	-(c_1+c_2)(\cP F\ga)\chi' - 2(3c_1-c_2)(\cP F\ga)\eta
\\\phan\non&&{}
	-\(\frac14+c_2\)(\ga F\Psi')
		-\(\frac34+2c_1-4c_2\)(\ga F\chi)
\\\phan\non&&{}
	+\frac12\(\frac14+c_1\)\hat
F\chi'+\(\frac14+c_1-2c_2\)\hat F\eta\biggr)
\\\phan\non&&{}
	+ 2c_2\(\bar\chi F\ga\)
		\(2\(\cP\cdot\Psi\)+(\cP\cdot\chi)+\chi' +2\eta\)
\\\phan\non&&{}
	- 4(c_1+c_2)\bar\eta\(\cP F\chi\)
	+ (1+4c_2)\bar\eta\(\ga F\chi\)
\\\phan\non&&{}
	- c_2\(\bar\chi' + 2\bar\eta\)(\hat \cP-1)\(\(\ga F\chi\) -2\hat
F\eta\)
\\\phan\non&&{}
	+ \(\(\bar\chi\cdot\cP\) + 2\bar\eta\)\(\(\ga F\chi\) -2\hat F\eta\)
\\\phan\non&&{}
	+\(\bar\chi' + 2\bar\eta\)\biggl(3c_1\(\cP F\chi\)
		+ \(\frac14+3c_2\)\(\ga F\chi\)
\\\phan\non&&{}
	 + 2(c_1-c_2)\(\cP F\ga\)\eta
		+ \(\frac14 + c_1\)\hat F\eta \biggr)
\\\phan&&{}
	- \(\(\bar\chi\cdot\cP\)+2\bar\eta\)\(\(\ga F\chi\)-2\hat F\eta\)
 + h.c.\biggr\}.
\eeqn
And from \reff{Trans0} we get the deformed gauge transformation for the
component fields
\beqn\label{Trans5/2}
\non
\de_1\Psi_{\mu\nu}&=& c_2(\cP F)_{(\mu}\xi_{\nu)}
					+\(\frac12+c_2\)(F\ga)_{(\mu}\xi_{\nu)},
\\\phan\non
\de_1\chi_\mu&=&\frac12c_2\cP_\mu(\ga F\xi)-\frac12c_2\hat F\cP_\mu\xi'
		+(c_1+c_2)(\ga F\cP)\xi_\mu + 3c_2(F\cP)_\mu\xi'
\\\phan\non&&{}
+(1+2c_1)(F\xi)_\mu
			- \frac12\(\frac14+c_1\)\hat F\xi_\mu
			- \(\frac14-3c_2\)(\ga F)_\mu\xi'
\\\phan\non
\de_1\eta &=& 2c_1(\cP F\xi) + (c_1-c_2)(\cP F\ga)\xi'
		-\frac32c_2(\ga F\xi)
\\\phan&&{}
		+\frac12\(\frac14+c_1-c_2\)\hat F\xi'.
\eeqn
In this, we have resolved the following relation with respect to $\veps$
\beq
\xi'-\frac12c_2(\ga F\xi) + \(1-\frac12c_2\hat F\)\veps + {\cal O}\(F^2\)
=0,
\eeq
which corresponds to \reff{xi'5/2} in linear approximation.

Having compared of the considered cases one can see that the case of the
massive spin-$\frac32$ field is rather marginal, because transformations
\reff{Trans3/2} can be reduced to the trivial ones by the redefinition of
field $\chi_\mu, \eta$ of type
\begin{eqnarray*}
\chi_\mu&\to& s_1\chi_\mu + s_2\ga_\mu\eta,
\\
\eta&\to& s_3\eta + s_4\chi',
\end{eqnarray*}
i.e. in the presence of the constant electromagnetic field all the
information about the non-minimal interaction can be transferred into the
Lagrangian. In this sense the situation for the spin-$\frac52$ and 
higher-spin fields is different, because from \reff{Trans5/2} it is clear 
that this transformation cannot be reduced to the trivial ones by any 
redefinitions of the fields.

\section{Conclusion}
Here we have extended the algebraic framework of the description of the 
massive bosonic fields with arbitrary spins to the case of the massive
fermionic ones. In such framework we have got the explicit form of the
operators by means of which we have obtained the Lagrangian and the gauge 
transformations describing the interaction between arbitrary massive 
half-integer fields and the constant electromagnetic field in linear 
approximation.  We have also applied our approach to the particular cases 
of the spin-$\frac32$ and spin-$\frac52$ fields and have derived the 
explicit gauge-inariant Lagrangian and the transformations for each case 
at linear order in the external e.m. field $F_{\mu\nu}$. 

Of course, there are open important questions: the existence of the next
approximation and the causality. We hope the following investigations will
shed light upon these questions.

It is worth noting that the case of the constant Abelian field can be
easily extended to that of a non-Abelian field. If so, we have to consider
the external field as the covariantly constant one. In this, one should take
the whole vacuum as $\ket{0}\otimes e^i$, where $e^i$ are basis vectors in
space of linear representation of a non-Abelian group. The covariant
derivative has the form $\partial_\mu+A_\mu^aT^a$, where $T^a$ are the
operators realizing the representation. Such modification does not change
the algebraic features of our scheme in linear approximation. Therefore, all
the results derived are valid in this case as well.

	In conclusion we would like to note that our approach allows one to
construct not only the electromagnetic interaction for the fields with
arbitrary spins, but it also allows one to describe the propagation of the
fields in a special Riemann space. For the integer spin fields such a
construction will be considered elsewere.

\section*{Acknowledgements}
\noindent
The author thanks Prof. Yu. M. Zinoviev for the useful discussion and the
help in the work.


\end{document}